\documentclass[twocolumn]{aastex62}

\usepackage{xspace}
\turnoffedit

\newcommand\msun {$M_{\odot}$\xspace}
\newcommand\rsun {$R_{\odot}$\xspace}

\def\lum{erg s$^{-1}$\xspace}

\def\chan{{\it Chandra}\xspace}
\def\hst{{\it HST}\xspace}

\def\gaia{{\it Gaia}\xspace}

\newcommand{\mhsta}{$m_\mathrm{F160W}$\xspace}
\newcommand{\mhstb}{$m_\mathrm{110W}$\xspace}
\newcommand{\mhstc}{$m_\mathrm{F814W}$\xspace}
\newcommand{\mhstd}{$m_\mathrm{F450W}$\xspace}
\newcommand{\mhste}{$m_\mathrm{F336W}$\xspace}
\newcommand{\mhstf}{$m_\mathrm{F225W}$\xspace}

\newcommand{\ngc}{NGC\,5907\xspace}
\newcommand{\ngculx}{NGC\,5907~X-1\xspace}
\newcommand{\ngculxt}{NGC\,5907~ULX-2\xspace}
\newcommand{\xone}{M82~X-1\xspace}
\newcommand{\xtwo}{M82~X-2\xspace}

\newcommand{\pth}{NGC\,7793~P13\xspace}
\newcommand{\snulx}{SN2010da\xspace}
\newcommand{\cyculx}{M51~ULX8\xspace}

\shorttitle{Donor stars of ULXPs}
\shortauthors{Heida et al.}

\begin{document}

\title{Searching for the Donor Stars of ULX Pulsars}

\correspondingauthor{M.~Heida}
\email{mheida@caltech.edu}

\author[0000-0002-1082-7496]{M.~Heida}
\affil{Cahill Center for Astronomy and Astrophysics, California Institute of Technology, 1200 E. California Blvd, Pasadena, CA 91125, USA}

\author{F.A.~Harrison}
\affil{Cahill Center for Astronomy and Astrophysics, California Institute of Technology, 1200 E. California Blvd, Pasadena, CA 91125, USA}

\author{M.~Brightman}
\affil{Cahill Center for Astronomy and Astrophysics, California Institute of Technology, 1200 E. California Blvd, Pasadena, CA 91125, USA}

\author{F.~F{\"u}rst}
\affil{European Space Astronomy Centre (ESAC), Science Operations Department, 28692 Villanueva de la Ca\~nada, Madrid, Spain}

\author[0000-0003-2686-9241]{D.~Stern}
\affil{Jet Propulsion Laboratory, California Institute of Technology, 4800 Oak Grove Drive, Mail Stop 169-221, Pasadena, CA 91109, USA}

\author{D.J.~Walton}
\affil{Institute of Astronomy, Madingley Road, Cambridge CB3 0HA, UK}



\begin{abstract}
We report on our search for the optical counterparts of two ultraluminous X-ray pulsars with known orbital periods, \xtwo and \ngculx, in new and archival \hst observations, in an effort to characterize the donor stars in these systems. We detect five near-infrared sources consistent with the position of \xtwo that are too bright to be single stars. We also detect \edit1{seven sources} in the WFC3/UVIS F336W image whose photometry matches that of \edit1{10 -- 15} \msun stars turning off the main sequence. Such stars have densities consistent with the properties of the donor star of \xtwo as inferred from X-ray timing analysis, although it is also possible that the donor is a lower mass star below our detection limit \edit1{or that there is a significant contribution from the accretion disc to the optical emission}. 
We detect three candidate counterparts to \ngculx in the near-infrared. All of these are too bright to be the donor star of the ULX, which based on its orbital period is a red giant. The high background at the location of \ngculx precludes us from detecting this expected donor star. The recently discovered \ngculxt also falls within the field of view of the near-infrared imaging; we detect \edit1{four} sources in the error circle, with photometry that matches AGB stars. The star suggested to be the counterpart of \ngculxt by \citet{pintore18} falls outside our 2-$\sigma$ error circle. 
\end{abstract}



\section{Introduction} \label{sec:intro}
Ultraluminous X-ray sources (ULXs) are off-nuclear point-like sources with $L_{\rm X} \geq 10^{39}$ \lum, exceeding the Eddington limit for a 10 \msun black hole \citep[e.g.,][]{fabbiano89}. Although early papers on ULXs suggested intermediate mass black holes (IMBHs; $10^{2} \lesssim M_{\rm{BH}} \lesssim10^{5}\,M_{\odot}$) as the accretors in these systems \citep[e.g.][]{colbert05}, a growing body of evidence --- mainly from broadband X-ray spectroscopy --- favors `stellar mass' objects exceeding the Eddington limit for the majority of sources \citep{gladstone09,sutton13,walton18}. A combination of super-Eddington accretion and geometric collimation would then explain the large luminosities. This interpretation was proven to be correct \edit1{for several sources} by the discovery of X-ray pulsations from \xtwo \citep{bachetti14} and, more recently, from three other ULXs \edit2{\citep[\ngculx, \pth and \snulx, also designated NGC\,300~ULX1;][]{fuerst16,israel17a,israel17b,carpano18,binder18}}, showing that in these objects the accretor is in fact a neutron star. \edit2{X-ray luminosities exceeding $10^{39}$ \lum have also been observed in outbursts from Be X-ray binaries with neutron star accretors \citep[e.g.~SMC X-3 and Swift J0243.6+6124,][]{tsygankov17,jaisawal18,wilsonhodge18}.} A fifth neutron star ULX (\cyculx), without X-ray pulses but identified through a cyclotron resonance feature, was reported by \citet{brightman18}. For a recent review on ULXs we refer the reader to \citet{kaaret17}.

Accreting magnetized neutron stars can, in principle, reach these super-Eddington luminosities through a number of mechanisms. For example, magnetic fields $\gtrsim 10^{12}$ G will collimate the accretion flow, allowing material to accrete onto the polar regions while radiation escapes from the sides of the column. Strong surface magnetic fields ($\gtrsim 10^{13.5}$ G) reduce the scattering cross section for electrons, reducing the radiation pressure and increasing the effective Eddington luminosity \citep{basko76,herold79,mushtukov15}. However, configurations with low magnetic fields have also been proposed \citep[e.g.][]{kluzniak15}.
All accretion theories trying to explain the seemingly super-Eddington luminosities of ULXs --- with either neutron star or stellar mass black hole accretors --- include a highly super-Eddington mass transfer rate from the donor star, which is difficult to reach through wind accretion. The favored scenario is therefore mass transfer through Roche lobe overflow from a massive donor star.

Very few ULX mass donor stars have been identified. The large distances to ULXs in combination with their often bright accretion discs make their donor stars hard to detect, and even more difficult to confirm spectroscopically. The only systems where spectroscopic signatures of donor stars are detected are M101 ULX1, with a Wolf-Rayet donor star \citep{liu13}; three ULXs (NGC 253 J004722.4-252051, NGC 925 J022721+333500 and NGC 4136 J120922+295559) with M-type supergiant donors \citep{heida15a,heida16}; and the ULX pulsar (ULXP) \pth, which has a blue supergiant (B9Ia) donor \citep{motch11,motch14}. The recently discovered ULXP SN2010da in NGC 300 likely has a supergiant (sg) B[e] or yellow sg donor star \citep{lau16,villar16}. These are all massive stars, confirming the idea that many ULXs are an extreme class of high-mass X-ray binaries and possible progenitors of gravitational wave sources \edit1{\citep{esposito15,inoue16,marchant17}}. 
However, there is a significant observational bias in favor of massive companions because they are much brighter than low-mass stars. \citet{wiktorowicz17} predict that the majority of neutron star ULXs have low mass ($\leq 1.5$ \msun), red giant donors.

For ULXs with a black hole accretor, phase-resolved spectroscopic observations of the donor star are the only direct way to obtain limits on the mass of the compact object. In ULXPs we can turn this around: since neutron stars have been observed to have a very narrow range of masses \citep[$1 - 2$ \msun, see][for a review]{lattimer12rev}, the mass of the compact object is known , and --- unless the system is viewed at a very low inclination --- we can obtain orbital parameters of the system by observing the modulation of the pulse period due to the Doppler effect. This gives us the mass function and therefore limits on the mass and average density of the donor star. In combination with direct observations of the donor it is then possible to investigate if these systems are indeed fed through Roche lobe overflow, as was recently shown for \pth \citep{fuerst18}. It is also a potential way to identify ULXs with low-mass donor stars.


In this paper we analyze new and archival {\it Hubble Space Telescope} (\hst) observations of \xtwo and \ngculx. We combine the photometry with orbital parameters of these systems derived from X-ray timing analyses by \citet{bachetti14} and \citet{israel17a} to put constraints on their respective donor stars. All magnitudes in this paper are Vega magnitudes.

\section{Data reduction and analysis} \label{sec:obs}
\subsection{\hst observations}
\begin{deluxetable*}{llllll}
\tablecaption{List of \hst observations \label{tab:obshst}}
\tablecolumns{6}
\tablewidth{0pt}
\tablehead{
\colhead{Target} & \colhead{Prop ID} & \colhead{Obs date} & \colhead{Inst.} & \colhead{Filter} & \colhead{Exp. time (s)}
}
\startdata
\ngc & 15074 & 2017-12-13 & WFC3/IR & F160W & 5612 \\
 & 6092 & 1996-03-31 & WFPC2 & F450W & 780 \\
 & 6092 & 1996-03-31 & WFPC2 & F814W &  480 \\
M82 & 11360 & 2010-01-01 & WFC3/UVIS & F225W & 1665 \\
 & 11360 & 2010-01-01 & WFC3/UVIS & F336W & 1620 \\
 & 11360 & 2010-01-01 & WFC3/IR & F110W & 1195 \\
 & 11360 & 2010-01-01 & WFC3/IR & F160W & 2395 \\
\enddata
\end{deluxetable*}

We obtained a deep \hst WFC3/IR F160W image of \ngc (proposal ID 15074) and downloaded archival WFPC2 images of \ngc \edit1{\citep{kissler-patig99}} and WFC3 images of M82 (\citealt[e.g.][]{lim13}; see Table \ref{tab:obshst}). There are additional deep WFPC2/F606W images of \ngc, but \ngculx is just outside the field of view. \edit2{The WFPC2/F450W and F814W observations of \ngc have previously been analyzed by \citet{sutton12} who initially reported a candidate counterpart to \ngculx that is only detected in the F450W image (\mhstd = $21.5 \pm 0.4$). However, on closer inspection, this source appeared to be spurious: it is only present in two of the three exposures with the F450W filter and likely due to cosmic rays that unfortunately hit at the same location in these two exposures \citep{sutton13b}.} \edit1{The observations of M82 were used by \citet{voss11} and \citet{wang15} to search for counterparts to M82 X-1, but not \xtwo. \citet{gladstone13} did search for counterparts to \xtwo in older \hst data (they call the source NGC~3034 ULX4) but they did not detect any candidate counterparts.}

We use DOLPHOT version 2.0 \citep{dolphin00} for the photometric analysis. Following the DOLPHOT manuals for the WFPC2 and WFC3 modules, we photometer the {\it flt} (WFC3) and {\it c0m} (WFPC2) images separately, using the drizzled images produced by the \hst pipeline as the reference image. The ULXs are located in crowded regions with high and variable backgrounds, especially \ngculx. We find that setting {\it fitsky = 3}, with other parameter values as recommended in the manuals, yields the best results. \edit1{To select good stars we filter on signal-to-noise (S/N; must be $> 5$), sharpness (between $-0.3$ and $+0.3$), object type (only stars with type 1, or `good stars') and photometry flags (only flag values of 0 are selected).}

\subsection{ULX positions}
\begin{deluxetable*}{lllll}
\tablecaption{List of \chan observations \label{tab:obschan}}
\tablecolumns{5}
\tablewidth{0pt}
\tablehead{
\colhead{Target} & \colhead{Obs ID} & \colhead{Obs start date} & \colhead{Inst.} & \colhead{Exp. time (ks)}
}
\startdata
\ngc & 12987 & 2012-02-11 & ACIS-S & 16.0 \\
 & 20830 & 2017-11-07 & ACIS-S & 51.3 \\
M82 & 10542 & 2009-06-24 & ACIS-S & 118.6 \\
 & 10543 & 2009-07-01 & ACIS-S & 118.5 \\
\enddata
\end{deluxetable*}

We determined the positions of the ULXs and other X-ray sources in archival deep \chan observations of the two galaxies (see Table \ref{tab:obschan}) using the {\sc CIAO} tool {\it wavdetect}. There are two very deep \chan observations covering \xtwo, but none of the X-ray sources have unique optical counterparts in the \hst images. Instead we register both \chan observations and the drizzled \hst images using \gaia DR2 sources in the field of view. We use the {\sc CIAO} tool {\it reproject\_aspect} to register the \chan observations, only considering X-ray sources with S/N $> 5$ (194 sources in obsid 10542 and 209 sources in obsid 10543, excluding \xtwo itself). There are 4 and 3 sources with matches in the \gaia catalog for obsid 10542 and 10543, respectively. Given the low number of sources we only calculate the x- and y-shift, leaving the rotation and scale fixed. We adopt the average residual distance between the \gaia and \chan positions, $0.3''$, as a measure of the astrometric uncertainty. The positional uncertainty due to localizing the ULX in the \chan image is negligible ($<0.05''$). The position of \xtwo in the two \chan observations is consistent to within $<0.1''$ --- we adopt the position of the ULX in obsid 10543, RA = 09:55:51.21, Dec = +69:40:44.1 (J2000). 
To register the drizzled \hst images we use the Starlink/GAIA `fit to star positions' tool, with 21 and 19 matched sources for the IR (F160W) and UVIS (F336W) images, respectively. We adopt the rms of the fit as the final astrometric uncertainty: this is $0.12''$ for the IR image and $0.04''$ for the UVIS image. The uncertainty of the ULX position on the \hst images is dominated by the uncertainty in the registration of the \chan observation and is $0.3''$ for both the IR and UVIS images.

For \ngc we retrieved two \chan observations from the archive. The shorter one (obsid 12987) contains \ngculx but no X-ray sources that can be used for cross-matching with the \hst image or other optical catalogs. The longer observation (obsid 20830) is a recent DDT observation obtained while \ngculx was in an off-state (\citealt{pintore18}; a previous off-state of this source was observed by \citealt{walton15b}). This observation does contain two X-ray sources with unique counterparts in our WFC3/IR image. Using {\it reproject\_aspect}, we first register the short \chan observation to the longer one using 8 common X-ray sources with S/N $> 4$. The average residual after the shift is $0.23''$. We then use the two overlapping sources between the longer \chan observation and our WFC3/IR image to register the \chan observations to the \hst image. With only two sources we cannot calculate the uncertainty in this step; we conservatively assume an error of $0.3''$. Quadratically adding the two uncertainties gives a total astrometric uncertainty of $0.4''$ (again, the error in the localization of the ULX in the \chan image is negligible). The position of the ULXP on the WFC3 image is RA = 15:15:58.66, Dec = +56:18:10.3 (J2000). 

The new ULX reported by \citet[][\ngculxt]{pintore18} is visible in \chan observation 20830. Its position in our WFC3 F160W image is RA = 15:16:01.13, Dec. = +56:17:51.5 (J2000), with an astrometric uncertainty of $0.3''$ that is mainly due to the registration of \chan observation 20830 to our \hst image. 

\subsection{Limiting magnitudes}
\edit1{The limiting magnitudes of the \hst observations vary across the images due to the local background. We use the {\sc Dolphot} {\it fakestar} routine to determine the limiting magnitudes at the positions of the ULXs. In every image we insert 2000 fake stars distributed normally around the position of the ULX, spanning a range of 8 input magnitudes. We then run {\sc Dolphot} in fakestar mode to obtain the photometry of these fake stars. We adopt as the limiting magnitude the faintest magnitude at which $\geq 90\%$ of the fake stars are retrieved with S/N $\geq 5$. To make sure our photometry is accurate, we also check that the retrieved magnitudes of the fake stars are consistent with their input magnitudes to within $0.3$ mag. In two cases (the WFC3/F160W and WFPC2/F814W observations of \ngculx) we find that the retrieved magnitudes diverge significantly from the input magnitudes. This is likely due to imperfect modeling of the complex background by {\sc Dolphot}. In these two cases we adopt the faintest magnitude at which $\geq 90\%$ of the fake stars are retrieved within 0.3 mag from their input magnitude as our limiting magnitude. All magnitude limits are listed in Table \ref{tab:limmag}; we only consider sources brighter than this limit in our analysis.}

\begin{deluxetable}{lll}
\tablecaption{Limiting magnitudes of \hst images at the position of the ULXs \label{tab:limmag}} 
\tablecolumns{3}
\tablewidth{0pt}
\tablehead{
\colhead{Source} & \colhead{\hst filter} & \colhead{Limiting magnitude}
}
\startdata
\xtwo & F225W & 25.1 \\
 & F336W & 26.4 \\
 & F110W & 18.8 \\
 & F160W & 18.8 \\
\ngculx & F450W & 25.5 \\
 & F814W & 23.6 \\
 & F160W & 22.0 \\
\ngculxt & F450W & 25.8 \\
 & F814W & 25.0 \\
 & F160W & 23.7 \\
\enddata
\end{deluxetable} 

\section{Results}
\subsection{\ngc}
\begin{figure*}
\includegraphics[width=\textwidth]{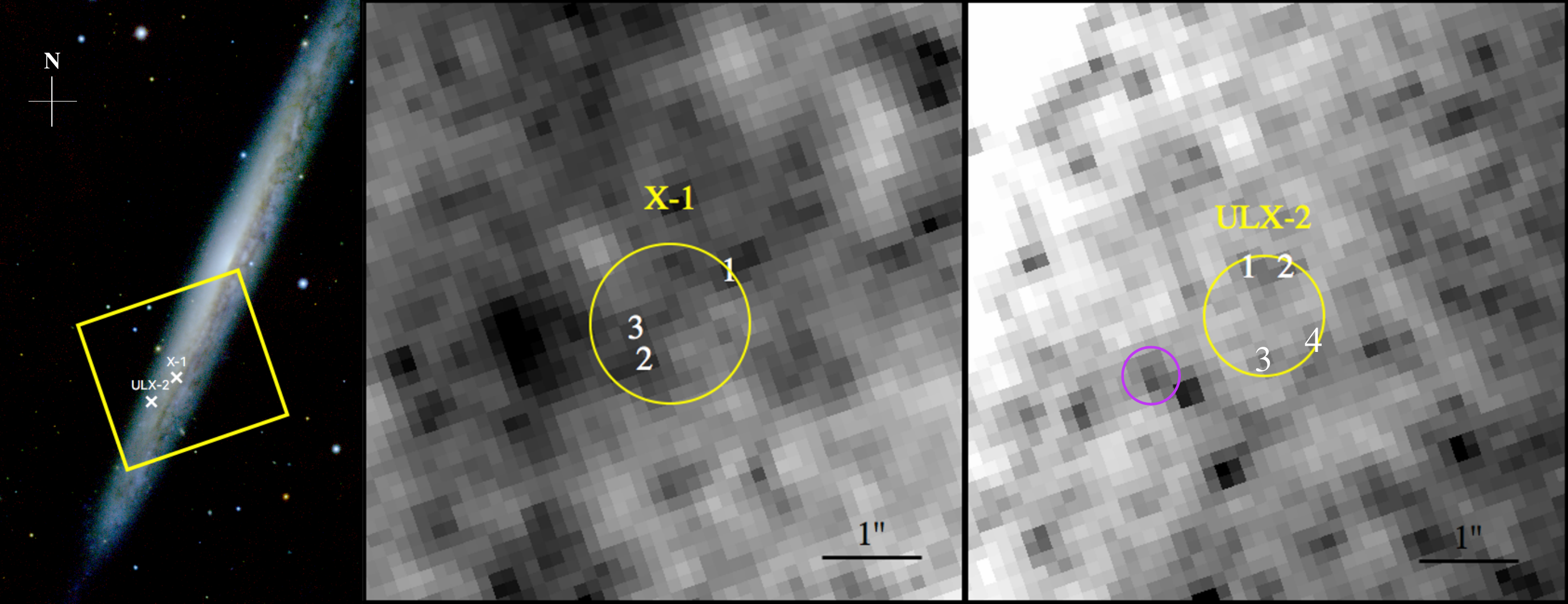}
\caption{{\it Left:} False-color PanSTARRS image of \ngc ({\it g'}, {\it i'} and {\it y'} bands) with the locations of the two ULXs marked with white crosses. The yellow box indicates the field of view of our WFC3/F160W observation. {\it Middle and right:} The locations of the two ULXs on our WFC3/F160W observations. The yellow circles indicate the 2-$\sigma$ regions around the X-ray positions of the ULXs. The retrieved magnitudes of the sources labeled in this Figure are listed in Table \ref{tab:ngcmag}. The counterpart to \ngculxt proposed by \citet{pintore18} is indicated with a magenta circle.} 
\label{fig:ngchst} 
\end{figure*}

We detect three sources brighter than the limiting magnitude inside the 2-$\sigma$ error circle of \ngculx in our F160W image (see center right panel in Figure \ref{fig:ngchst} and Table \ref{tab:ngcmag}). The errors listed are the statistical errors as calculated by {\sc Dolphot}. The brightest source is also detected in the F814W \edit1{and F450W} images, at \mhstc $\approx 23.5$ and \mhstd $\approx 25.0$. No \edit1{other} point sources are detected in the F450W \edit1{and F814W} images.

We detect \edit1{four} sources brighter than the limiting magnitude in the 2-$\sigma$ error circle of \ngculxt in our F160W image (see right-hand panel in Figure \ref{fig:ngchst} and Table \ref{tab:ngcmag}). \edit1{Only the brightest of these sources is detected in the F450W image. No sources are significantly detected in the F814W image.} The \citet{pintore18} preferred counterpart is visible in our F160W and F814W images (indicated with a magenta circle in Figure \ref{fig:ngchst}), but its position is inconsistent (at $> 3$-$\sigma$) with our localization of the ULX. This is due to our smaller localization uncertainty: our $1-\sigma$ uncertainty is $0.3''$ versus $0.42''$ for \citet{pintore18}, as they used a different method to register the \chan and \hst images.

\begin{deluxetable}{llll}
\tablecaption{{\sc Dolphot} magnitudes of sources labeled in Figure \ref{fig:ngchst}. Listed magnitudes are not corrected for extinction \label{tab:ngcmag}} 
\tablecolumns{4}
\tablewidth{0pt}
\tablehead{
\colhead{Source} & \colhead{\mhsta} & \colhead{\mhstc} & \colhead{\mhstd}
}
\startdata
\ngculx & & & \\
1& $21.61 \pm 0.01$ & $23.53 \pm 0.14$ & $25.0 \pm 0.2$ \\
2 & $21.75 \pm 0.01$ & $> 23.6$ & $> 25.5$ \\
3 & $21.91 \pm 0.02$ & $> 23.6$ & $> 25.5$ \\
\ngculxt & & & \\
1 & $22.68 \pm 0.02$ & $>25.0$ & $24.7 \pm 0.3$ \\
2 & $22.73 \pm 0.02$ & $>25.0$ & $> 25.8$ \\
3 & $23.21 \pm 0.03$ & $>25.0$ & $> 25.8$ \\
4 & $23.38 \pm 0.04$ & $>25.0$ & $> 25.8$ \\
 \enddata
\end{deluxetable} 

\subsection{\xtwo}
\begin{figure*}
\includegraphics[width=\textwidth]{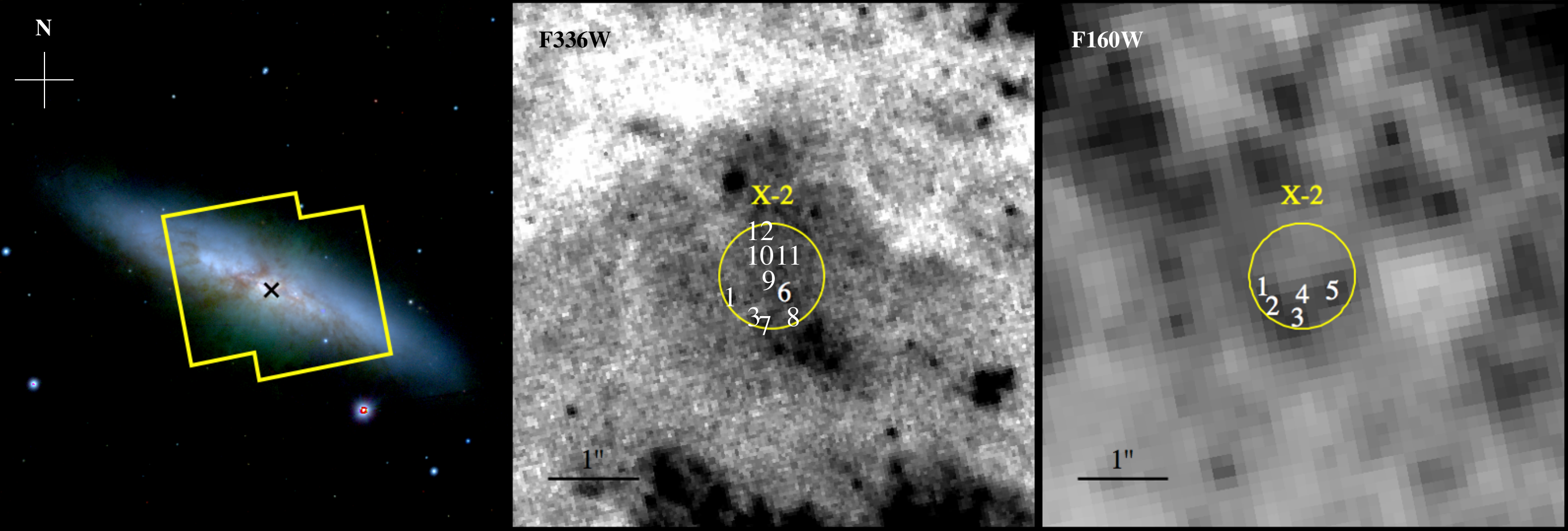}
\caption{{\it Left:} False-color PanSTARRS image of M82 ({\it g'}, {\it r'} and {\it y'} bands) with the location of \xtwo marked with a black cross. The yellow box indicates the field of view of the archival WFC3/NIR observations. The field of view of the WFC3/UVIS observations is similar. {\it Middle and right:} The location of \xtwo on the WFC3/F336W (middle) and WFC3/F160W (right) observations. The yellow circles indicate the 2-$\sigma$ region around the X-ray position of the ULX. The retrieved magnitudes of the sources labeled in this Figure are listed in Table \ref{tab:mmag}.} 
\label{fig:mhst}
\end{figure*}

We detect five sources in the 2-$\sigma$ error circle of \xtwo in the NIR images, with $17.4 \leq m_\mathit{F110W} \leq 18.4$ and $16.0 \leq m_\mathit{F160W} \leq 16.8$ (see right-hand panel in Figure \ref{fig:mhst} and Table \ref{tab:mmag}). \edit1{Two of these, as well as seven additional sources that are not detected in the NIR images, are detected in the F336W image (center panel in Figure \ref{fig:mhst}).} No point sources are significantly detected in the F225W image.

\begin{deluxetable}{lllll}
\tablecaption{{\sc Dolphot} magnitudes of sources labeled in Figure \ref{fig:mhst}. Listed magnitudes are not corrected for extinction \label{tab:mmag}} 
\tablecolumns{4}
\tablewidth{0pt}
\tablehead{
\colhead{Source} & \colhead{\mhsta} & \colhead{\mhstb} & \colhead{\mhste} & \colhead{\mhstf}
}
\startdata
\xtwo & & & & \\
1& $16.719 \pm 0.001$ & $18.411 \pm 0.002$ & $24.9 \pm 0.1$ & $>25.1$ \\
2 & $16.006 \pm 0.001$ & $17.386 \pm 0.001$ & $> 26.4$ & $>25.1$ \\
3 & $16.153 \pm 0.001$ & $17.537 \pm 0.001$ & $25.3 \pm 0.2$ & $>25.1$ \\
4 & $16.822 \pm 0.002$ & $18.442 \pm 0.002$ & $> 26.4$ & $>25.1$ \\
5 & $16.046 \pm 0.001$ & $17.436 \pm 0.001$ & $> 26.4$ & $>25.1$ \\
6 & $> 18.8$ & $> 18.8$ & $24.39 \pm 0.09$ & $>25.1$ \\
7 & $> 18.8$ & $> 18.8$ & $25.4 \pm 0.2$ & $>25.1$ \\
8 & $> 18.8$ & $> 18.8$ & $25.3 \pm 0.2$ & $>25.1$ \\
9 & $> 18.8$ & $> 18.8$ & $25.6 \pm 0.2$ & $>25.1$ \\
10 & $> 18.8$ & $> 18.8$ & $25.6 \pm 0.2$ & $>25.1$ \\
11 & $> 18.8$ & $> 18.8$ & $25.9 \pm 0.3$ & $>25.1$ \\
12 & $> 18.8$ & $> 18.8$ & $25.7 \pm 0.3$ & $>25.1$ \\
 \enddata
\end{deluxetable} 

\section{Discussion} \label{sec:disc}
We searched for counterparts to two ULXPs in new and archival \hst observations. Both sources are located in crowded regions with high extinction --- \xtwo in the center of M82, and \ngculx in the dust lane of the edge-on spiral galaxy \ngc. We detect multiple potential counterparts to both ULXPs as well as to a second, recently discovered ULX in \ngc. 
Given the distance modulus to \ngc ($m - M = 31.17 \pm 0.09$, \edit1{corresponding to a distance of 17.1 Mpc;} \citealt{2013AJ....146...86T}), the absolute magnitudes of the counterparts to \ngculx in the F160W filter are $-9.5 \leq M_\mathit{F160W} \leq -9.3$. \edit1{Counterpart 1 is also detected in the F450W and F814W filters with absolute magnitudes of $M_\mathit{F814W} \approx -7.6$ and $M_\mathit{F450W} \approx -6.0$, while the other two counterparts are fainter than $-7.6$ mag in the F814W filter and $-5.7$ mag in the F450W filter, respectively.} 
\edit1{The counterparts to \ngculxt have $-8.5 \leq M_\mathit{F160W} \leq -7.8$. Counterpart 1 is also detected in the F450W filter with absolute magnitude $M_\mathit{F450W} \approx -6.5$; the other counterparts are fainter than $-5.4$ mag in the F450W filter and all counterparts are fainter than $-6.2$ mag in the F814W filter.}

\edit1{The five counterparts of \xtwo (with a distance modulus of $27.74 \pm 0.08$, corresponding to a distance of 3.5 Mpc; \citealt{2013AJ....146...86T}) that are detected in the near-IR images have absolute magnitudes $-11.7 \leq M_\mathit{F160W} \leq -10.9$ and $-10.4 \leq M_\mathit{F110W} \leq -9.2$. Counterparts 1 and 3 are also detected in the F336W image with $M_\mathit{F336W} \approx -2.8$ and $-2.4$, respectively, while the other three are fainter than $-1.3$ mag in that band. The other seven counterparts are only detected in the F336W image, with $-3.4 \leq M_\mathit{F336W} \leq -1.8$. They are fainter than $-8.9$ mag in both the F110W and F160W bands. All counterparts are fainter than $-2.6$ mag in the F225W band.}

\subsection{Reddening}
\edit1{The foreground Galactic extinction in the direction of \ngc is negligible ($A_V \approx 0.03$, \citealt{schlegel98}). However, \ngculx is located on the dust lane of this edge-on spiral (see Fig.~\ref{fig:ngchst}) and the extinction due to dust and gas in the galaxy itself is significant.} \edit2{The hydrogen column density as measured in the X-ray spectra is $N_\mathrm{H} = 0.4 - 0.9 \times 10^{22}$ cm$^{-2}$ \citep{sutton13,walton15b,fuerst17}.} \edit1{This translates into a $V$-band absorption $A_V \approx 1.8$ -- $4.1$ \citep{guver09}. It is not clear from the X-ray observations whether the absorption is variable or not \citep{fuerst17}. If it is variable, part of this X-ray absorbing material is likely local to the system itself and may not obscure the companion star, and $A_V$ is likely closer to the lower end of this range.}

\ngculxt is located further away from the central dust lane; \citet{pintore18} found no local extinction on top of the foreground Galactic absorption from their analysis of the X-ray spectrum of this source.

The foreground Galactic extinction in the direction of M82 is $A_V = 0.485$ \citep{schlegel98}. On top of that there is significant extinction due to gas and dust ($A_V \geq 1.55$) in the core of the galaxy, where \xtwo is located \citep{hutton14}.  

\subsection{Limits on donor star properties from X-ray timing}
Assuming circular orbits and accretion via Roche lobe overflow, the density of the donor star can be determined directly given the orbital period of the binary system and the mass of the accretor, using the equation for the Roche radius \citep{eggleton83} and Kepler's laws. 
For \xtwo, the orbital period was reported by \citet{bachetti14} as $2.53260 \pm 0.00005$ days, with an eccentricity $\epsilon < 0.003$ and a projected semi-major axis ($a \sin(i)$) of $22.225 \pm 0.004$ ls.

For \ngculx these values are not as well constrained. \citet{israel17a} performed a likelihood analysis to obtain the orbital parameters of the system (their Figure 2). They found a most-probable period $P_\mathrm{orb} = 5.3^{+2.0}_{-0.9}$ days with $a \sin(i) = 2.5^{+4.3}_{-0.8}$ ls. 

For \xtwo, the lower limit on the donor star mass is $\sim 5$ \msun, assuming a neutron star mass of 1.4 \msun and a system inclination $\leq 60^\circ$ \citep{bachetti14}. For \ngculx, the mass function for the most likely parameters reported by \citet{israel17a} is $6 \times 10^{-4}$ \msun, implying a lower limit for the donor star of $\sim 0.1$ \msun. The measurements of \citet{israel17a} also allow for a longer orbital period (up to $\sim 21$ days) with a larger projected semi-major axis, that would imply a higher minimum mass for the donor star (e.g.~29 \msun for an orbital period of 18 days and $a \sin(i) = 200$ ls). 

\begin{figure}
\includegraphics[width=0.5\textwidth]{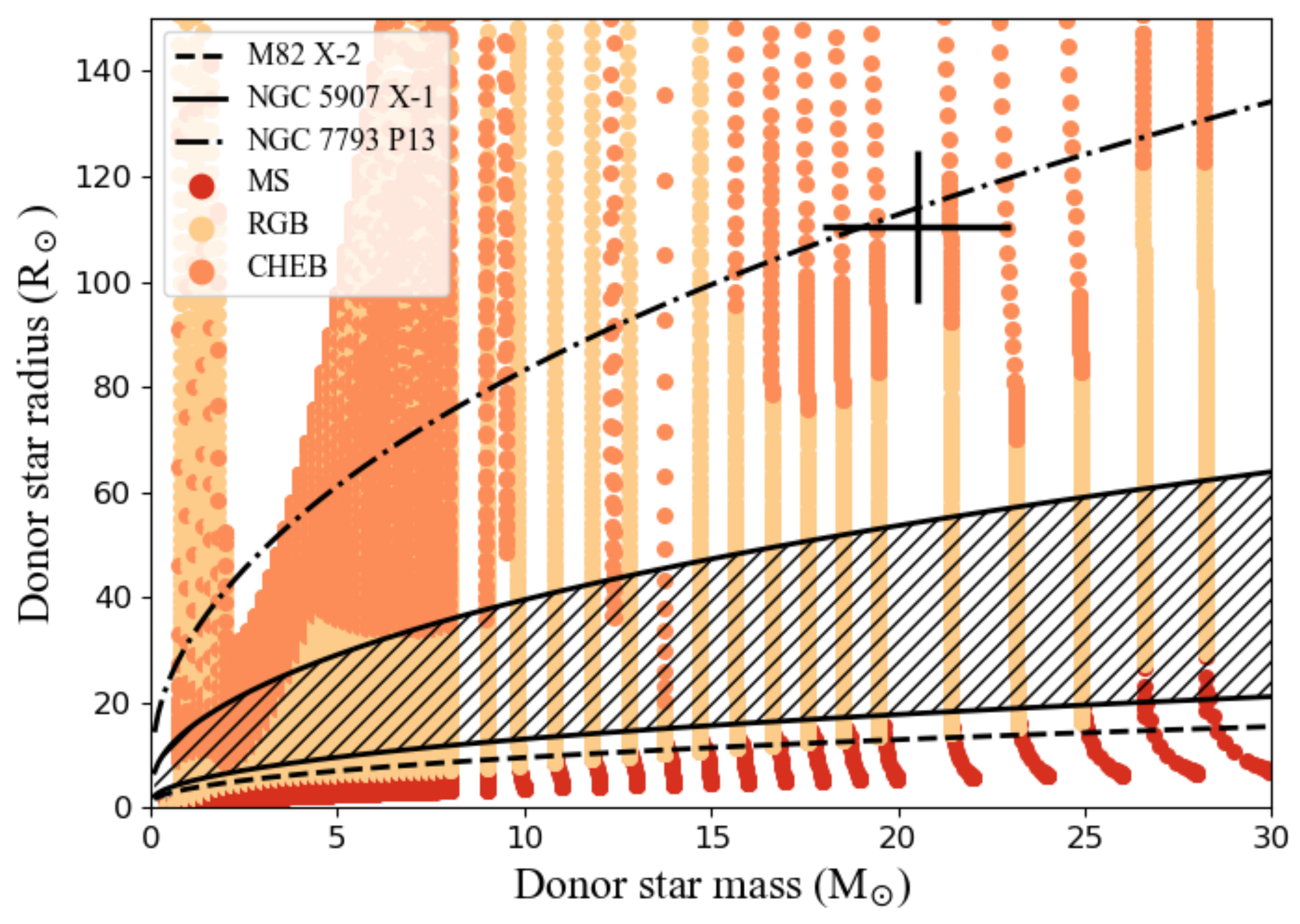}
\caption{Acceptable values of masses and radii of the donor stars of three ULXPs based on their orbital periods, assuming they accrete through Roche-lobe overflow. The lines plotted here assume a neutron star mass of 1.4 \msun, but are essentially the same for a neutron star mass of 2 \msun. The shaded region indicates the range of allowed orbital periods for \ngculx (4 - 21 days). The black cross indicates the mass and radius of the donor star of \pth \citep{motch14,fuerst18}. In the background are plotted evolution tracks of single stars at solar metallicity obtained from the MIST project. The evolutionary stages plotted here are main sequence (MS), red giant branch including the Hertzsprung gap (RGB) and core helium burning (CHEB).}
\label{fig:massrad}
\end{figure}

We plot the relation between allowed masses and radii for \xtwo and \ngculx in Figure \ref{fig:massrad}. For comparison, we also show the relation for \pth, a ULXP with a blue supergiant companion with a mass of 18 -- 23 \msun and a radius of 96 -- 125 \rsun \citep{motch14,fuerst18}. In the background we show evolution tracks of single, non-rotating stars at solar metallicity from the MESA Isochrones and Stellar Tracks (MIST) project \citep{choi16, dotter16}. \edit1{The donor stars in ULXPs obviously do not evolve in isolation, which will impact their evolution tracks \citep[see e.g.][]{rappaport05,patruno08,patruno10,ambrosi18}. However, comparing the possible masses and radii of the ULXP donors with those of single stars is useful to show the most plausible evolutionary stage of the donor stars in these systems.}

From Figure \ref{fig:massrad} we can see that the donor star of \xtwo is most likely just turning off the main sequence (as was also shown by \citealt{fragos15}), while the donor of \ngculx \edit1{should be a more evolved star in the Hertzsprung gap or on the red giant branch}.


\edit1{\subsection{Contribution from the accretion disc}
The colors and magnitudes of the optical counterparts of ULXs can be significantly affected by irradiation of the donor star and emission from the (supercritical) accretion disc (see e.g. \citealt{copperwheat05,copperwheat07,patruno08,patruno10,ambrosi18} for theoretical work, and \citealt{roberts11,grise12} for observations). The theoretical work has so far mostly focused on systems with a (massive) black hole as the accretor, and shows larger effects for more massive black holes; the impact on the optical counterparts of neutron star ULXs may be less pronounced. For example, the optical counterpart of \pth is affected by irradiation of the donor star, resulting in an observed spectral type that changes with the orbital phase of the system. However, its supergiant donor completely dominates the optical emission and there does not appear to be a significant contribution from an accretion disc to the optical light \citep{motch14}. Simulations like those of \citet{ambrosi18} for systems with a neutron star accretor would be very useful in assessing the effect of irradiation and the accretion disc on the optical/near-IR emission for ULXPs with non-supergiant donors. In the following we assume that the optical/near-IR emission from the ULXPs is dominated by the donor star as is the case for \pth.}

\edit1{\subsection{Limits on donor star properties from photometry}}
\edit1{To check if the sources we detect in the \hst images could be the donor stars of these ULXPs we downloaded synthetic WFC3 and WFPC2 photometry files at solar metallicity with $A_V = 1.8$ and $4.1$ (for \ngculx) and at [Fe/H] $= -0.25$ \citep{nagao11} and $A_V = 2.0$ (for \xtwo) from the MIST website, as well as WFC3 and WFPC2 photometry at solar metallicity with $A_V = 0.03$ for \ngculxt. In the case of \ngculx, all stars with magnitudes consistent with the observed ones are red supergiant stars with radii larger than 400 \rsun --- none of these can be the stellar companion to the ULXP.} The limiting magnitude of our WFC3/F160W image is too bright to detect the donor star, due to the high background in this region. The F160W magnitudes and F450W and F814W limits of the sources detected in the error circle of \ngculxt are consistent with asymptotic giant branch (AGB) stars. Without information on the orbital period and accretor mass in this system, we cannot say anything about the likelihood that one of these sources is the donor star to the ULX.

The near-IR counterparts to \xtwo are even brighter, and no single star in the MIST simulations matches these F110W and F160W absolute magnitudes. These sources \edit1{may be compact star clusters --- the resolution of the WFC3/IR camera at 1.6$\mu$m corresponds to $\sim 3$ pc at the distance of M82. The magnitudes and limits of sources that are only detected in the WFC3/UVIS F336W image do match those of single stars with masses and radii consistent with a possible donor of \xtwo: 10 -- 15 \msun stars just turning off the main sequence. One of these counterparts could potentially be the donor star of \xtwo, although it is also possible that the donor is a lower mass star too faint to be detected in the F336W image. In addition, one of these sources may be the real counterpart, but dominated by irradiation and/or emission from the accretion disc.} 

\section{Conclusions} \label{sec:con}
We obtained new WFC3/NIR images of \ngc and used those in conjunction with deep archival WFPC2 images to search for the donor star of \ngculx. The sources we detect are too bright to be the donor star of the ULXP, which based on its orbital period should be a red giant (see Figure \ref{fig:massrad}). The high background at the location of \ngculx precludes us from detecting the expected donor star, \edit1{assuming that the donor dominates the optical/near-IR emission from the system}. The recently discovered \ngculxt also falls within the field of view; we detect \edit1{four} sources in the error circle, with photometry that matches AGB stars. The star suggested to be the counterpart by \citet{pintore18} falls outside our 2-$\sigma$ error circle. 

We also retrieved deep WFC3/NIR and WFC3/UVIS images of \xtwo from the \hst archive to search for the donor star of this ULXP. The sources detected in the NIR images are too bright to be single stars, but we also detect \edit1{several sources} in the UVIS F336W image whose photometry matches that of \edit1{10 -- 15} \msun stars turning off the main sequence. Such stars have densities consistent with the donor star of \xtwo, although it is also possible that the donor is a lower mass star fainter than our detection limit \edit1{or that the optical light is dominated by irradiation and/or emission from the accretion disc}. 

Characterizing the donor stars of ULXPs is important for testing models of binary evolution. To date, the only system with a spectroscopically characterized donor star is \pth, which hosts a blue supergiant star \citep{motch11, motch14}. However, the majority of ULXPs are expected to have lower mass red giant donor stars \citep{wiktorowicz17}. Shorter period systems such as \xtwo and \ngculx may be examples of such lower mass systems, but the crowded environments and the large distances to these sources make direct detection and especially spectroscopic characterization unfeasible with currently available instruments. For the nearest ULXs this may change when the first 30-40 m class telescopes become operational.

\acknowledgements
%
We would like to thank the anonymous referee whose comments helped to significantly improve the paper.

Based on observations made with the NASA/ESA Hubble Space Telescope, obtained from the Data Archive at the Space Telescope Science Institute, which is operated by the Association of Universities for Research in Astronomy, Inc., under NASA contract NAS 5-26555. These observations are associated with program 15074.
Support for Program number 15074 was provided by NASA through a grant from the Space Telescope Science Institute, which is operated by the Association of Universities for Research in Astronomy, Incorporated, under NASA contract NAS 5-26555.
The scientific results reported in this article are based in part on data obtained from the Chandra Data Archive.
This work has made use of data from the European Space Agency (ESA)
mission {\it Gaia} (\url{https://www.cosmos.esa.int/gaia}), processed by
the {\it Gaia} Data Processing and Analysis Consortium (DPAC,
\url{https://www.cosmos.esa.int/web/gaia/dpac/consortium}). Funding
for the DPAC has been provided by national institutions, in particular
the institutions participating in the {\it Gaia} Multilateral Agreement.
The work of DS was carried out at the Jet Propulsion Laboratory,
California Institute of Technology, under a contract with NASA.

\vspace{5mm}
\facilities{HST (WFPC2 and WFC3), Chandra (ACIS), Gaia}


\software{astropy \citep{2013A&A...558A..33A},  
DOLPHOT \citep{dolphin00},
MESA \citep{paxton11,paxton13,paxton15},
CIAO \citep{fruscione06}
          }

\bibliographystyle{aasjournal}
\bibliography{bibliography}

\begin{thebibliography}{}
\expandafter\ifx\csname natexlab\endcsname\relax\def\natexlab#1{#1}\fi
\providecommand{\url}[1]{\href{#1}{#1}}
\providecommand{\dodoi}[1]{doi:~\href{http://doi.org/#1}{\nolinkurl{#1}}}
\providecommand{\doeprint}[1]{\href{http://ascl.net/#1}{\nolinkurl{http://ascl.net/#1}}}
\providecommand{\doarXiv}[1]{\href{https://arxiv.org/abs/#1}{\nolinkurl{https://arxiv.org/abs/#1}}}

\bibitem[{{Ambrosi} \& {Zampieri}(2018)}]{ambrosi18}
{Ambrosi}, E., \& {Zampieri}, L. 2018, \mnras, 480, 4918,
  \dodoi{10.1093/mnras/sty2213}

\bibitem[{{Astropy Collaboration} {et~al.}(2013){Astropy Collaboration},
  {Robitaille}, {Tollerud}, {Greenfield}, {Droettboom}, {Bray}, {Aldcroft},
  {Davis}, {Ginsburg}, {Price-Whelan}, {Kerzendorf}, {Conley}, {Crighton},
  {Barbary}, {Muna}, {Ferguson}, {Grollier}, {Parikh}, {Nair}, {Unther},
  {Deil}, {Woillez}, {Conseil}, {Kramer}, {Turner}, {Singer}, {Fox}, {Weaver},
  {Zabalza}, {Edwards}, {Azalee Bostroem}, {Burke}, {Casey}, {Crawford},
  {Dencheva}, {Ely}, {Jenness}, {Labrie}, {Lim}, {Pierfederici}, {Pontzen},
  {Ptak}, {Refsdal}, {Servillat}, \& {Streicher}}]{2013A&A...558A..33A}
{Astropy Collaboration}, {Robitaille}, T.~P., {Tollerud}, E.~J., {et~al.} 2013,
  \aap, 558, A33, \dodoi{10.1051/0004-6361/201322068}

\bibitem[{{Bachetti} {et~al.}(2014){Bachetti}, {Harrison}, {Walton},
  {Grefenstette}, {Chakrabarty}, {F{\"u}rst}, {Barret}, {Beloborodov}, {Boggs},
  {Christensen}, {Craig}, {Fabian}, {Hailey}, {Hornschemeier}, {Kaspi},
  {Kulkarni}, {Maccarone}, {Miller}, {Rana}, {Stern}, {Tendulkar}, {Tomsick},
  {Webb}, \& {Zhang}}]{bachetti14}
{Bachetti}, M., {Harrison}, F.~A., {Walton}, D.~J., {et~al.} 2014, \nat, 514,
  202, \dodoi{10.1038/nature13791}

\bibitem[{{Basko} \& {Sunyaev}(1976)}]{basko76}
{Basko}, M.~M., \& {Sunyaev}, R.~A. 1976, \mnras, 175, 395,
  \dodoi{10.1093/mnras/175.2.395}

\bibitem[{{Binder} {et~al.}(2018){Binder}, {Levesque}, \&
  {Dorn-Wallenstein}}]{binder18}
{Binder}, B., {Levesque}, E.~M., \& {Dorn-Wallenstein}, T. 2018, \apj, 863,
  141, \dodoi{10.3847/1538-4357/aad3bd}

\bibitem[{{Brightman} {et~al.}(2018){Brightman}, {Harrison}, {F{\"u}rst},
  {Middleton}, {Walton}, {Stern}, {Fabian}, {Heida}, {Barret}, \&
  {Bachetti}}]{brightman18}
{Brightman}, M., {Harrison}, F.~A., {F{\"u}rst}, F., {et~al.} 2018, Nature
  Astronomy, 2, 312, \dodoi{10.1038/s41550-018-0391-6}

\bibitem[{{Carpano} {et~al.}(2018){Carpano}, {Haberl}, {Maitra}, \&
  {Vasilopoulos}}]{carpano18}
{Carpano}, S., {Haberl}, F., {Maitra}, C., \& {Vasilopoulos}, G. 2018, \mnras,
  476, L45, \dodoi{10.1093/mnrasl/sly030}

\bibitem[{{Choi} {et~al.}(2016){Choi}, {Dotter}, {Conroy}, {Cantiello},
  {Paxton}, \& {Johnson}}]{choi16}
{Choi}, J., {Dotter}, A., {Conroy}, C., {et~al.} 2016, \apj, 823, 102,
  \dodoi{10.3847/0004-637X/823/2/102}

\bibitem[{{Colbert} \& {Miller}(2005)}]{colbert05}
{Colbert}, E.~J.~M., \& {Miller}, M.~C. 2005, in The Tenth Marcel Grossmann
  Meeting. On recent developments in theoretical and experimental general
  relativity, gravitation and relativistic field theories, ed. {M.~Novello,
  S.~Perez Bergliaffa, \& R.~Ruffini}, 530--+

\bibitem[{{Copperwheat} {et~al.}(2005){Copperwheat}, {Cropper}, {Soria}, \&
  {Wu}}]{copperwheat05}
{Copperwheat}, C., {Cropper}, M., {Soria}, R., \& {Wu}, K. 2005, \mnras, 362,
  79, \dodoi{10.1111/j.1365-2966.2005.09223.x}

\bibitem[{{Copperwheat} {et~al.}(2007){Copperwheat}, {Cropper}, {Soria}, \&
  {Wu}}]{copperwheat07}
---. 2007, \mnras, 376, 1407, \dodoi{10.1111/j.1365-2966.2007.11551.x}

\bibitem[{{Dolphin}(2000)}]{dolphin00}
{Dolphin}, A.~E. 2000, \pasp, 112, 1383, \dodoi{10.1086/316630}

\bibitem[{{Dotter}(2016)}]{dotter16}
{Dotter}, A. 2016, \apjs, 222, 8, \dodoi{10.3847/0067-0049/222/1/8}

\bibitem[{{Eggleton}(1983)}]{eggleton83}
{Eggleton}, P.~P. 1983, \apj, 268, 368, \dodoi{10.1086/160960}

\bibitem[{{Esposito} {et~al.}(2015){Esposito}, {Israel}, {Milisavljevic},
  {Mapelli}, {Zampieri}, {Sidoli}, {Fabbiano}, \& {Rodr{\'{\i}}guez
  Castillo}}]{esposito15}
{Esposito}, P., {Israel}, G.~L., {Milisavljevic}, D., {et~al.} 2015, \mnras,
  452, 1112, \dodoi{10.1093/mnras/stv1379}

\bibitem[{{Fabbiano}(1989)}]{fabbiano89}
{Fabbiano}, G. 1989, \araa, 27, 87, \dodoi{10.1146/annurev.aa.27.090189.000511}

\bibitem[{{Fragos} {et~al.}(2015){Fragos}, {Linden}, {Kalogera}, \&
  {Sklias}}]{fragos15}
{Fragos}, T., {Linden}, T., {Kalogera}, V., \& {Sklias}, P. 2015, \apj, 802,
  L5, \dodoi{10.1088/2041-8205/802/1/L5}

\bibitem[{{Fruscione} {et~al.}(2006){Fruscione}, {McDowell}, {Allen},
  {Brickhouse}, {Burke}, {Davis}, {Durham}, {Elvis}, {Galle}, {Harris},
  {Huenemoerder}, {Houck}, {Ishibashi}, {Karovska}, {Nicastro}, {Noble},
  {Nowak}, {Primini}, {Siemiginowska}, {Smith}, \& {Wise}}]{fruscione06}
{Fruscione}, A., {McDowell}, J.~C., {Allen}, G.~E., {et~al.} 2006, in Society
  of Photo-Optical Instrumentation Engineers (SPIE) Conference Series, Vol.
  6270, Society of Photo-Optical Instrumentation Engineers (SPIE) Conference
  Series

\bibitem[{{F{\"u}rst} {et~al.}(2017){F{\"u}rst}, {Walton}, {Stern}, {Bachetti},
  {Barret}, {Brightman}, {Harrison}, \& {Rana}}]{fuerst17}
{F{\"u}rst}, F., {Walton}, D.~J., {Stern}, D., {et~al.} 2017, \apj, 834, 77,
  \dodoi{10.3847/1538-4357/834/1/77}

\bibitem[{{F{\"u}rst} {et~al.}(2016){F{\"u}rst}, {Walton}, {Harrison}, {Stern},
  {Barret}, {Brightman}, {Fabian}, {Grefenstette}, {Madsen}, {Middleton},
  {Miller}, {Pottschmidt}, {Ptak}, {Rana}, \& {Webb}}]{fuerst16}
{F{\"u}rst}, F., {Walton}, D.~J., {Harrison}, F.~A., {et~al.} 2016, \apjl, 831,
  L14, \dodoi{10.3847/2041-8205/831/2/L14}

\bibitem[{{F{\"u}rst} {et~al.}(2018){F{\"u}rst}, {Walton}, {Heida}, {Harrison},
  {Barret}, {Brightman}, {Fabian}, {Middleton}, {Pinto}, {Rana}, {Tramper},
  {Webb}, \& {Kretschmar}}]{fuerst18}
{F{\"u}rst}, F., {Walton}, D.~J., {Heida}, M., {et~al.} 2018, \aap, 616, A186,
  \dodoi{10.1051/0004-6361/201833292}

\bibitem[{{Gladstone} {et~al.}(2013){Gladstone}, {Copperwheat}, {Heinke},
  {Roberts}, {Cartwright}, {Levan}, \& {Goad}}]{gladstone13}
{Gladstone}, J.~C., {Copperwheat}, C., {Heinke}, C.~O., {et~al.} 2013, \apjs,
  206, 14, \dodoi{10.1088/0067-0049/206/2/14}

\bibitem[{{Gladstone} {et~al.}(2009){Gladstone}, {Roberts}, \&
  {Done}}]{gladstone09}
{Gladstone}, J.~C., {Roberts}, T.~P., \& {Done}, C. 2009, \mnras, 397, 1836,
  \dodoi{10.1111/j.1365-2966.2009.15123.x}

\bibitem[{{Gris{\'e}} {et~al.}(2012){Gris{\'e}}, {Kaaret}, {Corbel}, {Feng},
  {Cseh}, \& {Tao}}]{grise12}
{Gris{\'e}}, F., {Kaaret}, P., {Corbel}, S., {et~al.} 2012, \apj, 745, 123,
  \dodoi{10.1088/0004-637X/745/2/123}

\bibitem[{{G{\"u}ver} \& {{\"O}zel}(2009)}]{guver09}
{G{\"u}ver}, T., \& {{\"O}zel}, F. 2009, \mnras, 400, 2050,
  \dodoi{10.1111/j.1365-2966.2009.15598.x}

\bibitem[{{Heida} {et~al.}(2016){Heida}, {Jonker}, {Torres}, {Roberts},
  {Walton}, {Moon}, {Stern}, \& {Harrison}}]{heida16}
{Heida}, M., {Jonker}, P.~G., {Torres}, M.~A.~P., {et~al.} 2016, \mnras, 459,
  771, \dodoi{10.1093/mnras/stw695}

\bibitem[{{Heida} {et~al.}(2015){Heida}, {Torres}, {Jonker}, {Servillat},
  {Repetto}, {Roberts}, {Walton}, {Moon}, \& {Harrison}}]{heida15a}
{Heida}, M., {Torres}, M.~A.~P., {Jonker}, P.~G., {et~al.} 2015, \mnras, 453,
  3510, \dodoi{10.1093/mnras/stv1853}

\bibitem[{{Herold}(1979)}]{herold79}
{Herold}, H. 1979, \prd, 19, 2868, \dodoi{10.1103/PhysRevD.19.2868}

\bibitem[{{Hutton} {et~al.}(2014){Hutton}, {Ferreras}, {Wu}, {Kuin},
  {Breeveld}, {Yershov}, {Cropper}, \& {Page}}]{hutton14}
{Hutton}, S., {Ferreras}, I., {Wu}, K., {et~al.} 2014, \mnras, 440, 150,
  \dodoi{10.1093/mnras/stu185}

\bibitem[{{Inoue} {et~al.}(2016){Inoue}, {Tanaka}, \& {Isobe}}]{inoue16}
{Inoue}, Y., {Tanaka}, Y.~T., \& {Isobe}, N. 2016, \mnras, 461, 4329,
  \dodoi{10.1093/mnras/stw1637}

\bibitem[{{Israel} {et~al.}(2017{\natexlab{a}}){Israel}, {Belfiore}, {Stella},
  {Esposito}, {Casella}, {De Luca}, {Marelli}, {Papitto}, {Perri}, {Puccetti},
  {Castillo}, {Salvetti}, {Tiengo}, {Zampieri}, {D'Agostino}, {Greiner},
  {Haberl}, {Novara}, {Salvaterra}, {Turolla}, {Watson}, {Wilms}, \&
  {Wolter}}]{israel17a}
{Israel}, G.~L., {Belfiore}, A., {Stella}, L., {et~al.} 2017{\natexlab{a}},
  Science, 355, 817, \dodoi{10.1126/science.aai8635}

\bibitem[{{Israel} {et~al.}(2017{\natexlab{b}}){Israel}, {Papitto}, {Esposito},
  {Stella}, {Zampieri}, {Belfiore}, {Rodr{\'\i}guez Castillo}, {De Luca},
  {Tiengo}, {Haberl}, {Greiner}, {Salvaterra}, {Sandrelli}, \&
  {Lisini}}]{israel17b}
{Israel}, G.~L., {Papitto}, A., {Esposito}, P., {et~al.} 2017{\natexlab{b}},
  \mnras, 466, L48, \dodoi{10.1093/mnrasl/slw218}

\bibitem[{{Jaisawal} {et~al.}(2018){Jaisawal}, {Naik}, \&
  {Chenevez}}]{jaisawal18}
{Jaisawal}, G.~K., {Naik}, S., \& {Chenevez}, J. 2018, \mnras, 474, 4432,
  \dodoi{10.1093/mnras/stx3082}

\bibitem[{{Kaaret} {et~al.}(2017){Kaaret}, {Feng}, \& {Roberts}}]{kaaret17}
{Kaaret}, P., {Feng}, H., \& {Roberts}, T.~P. 2017, \araa, 55, 303,
  \dodoi{10.1146/annurev-astro-091916-055259}

\bibitem[{{Kissler-Patig} {et~al.}(1999){Kissler-Patig}, {Ashman}, {Zepf}, \&
  {Freeman}}]{kissler-patig99}
{Kissler-Patig}, M., {Ashman}, K.~M., {Zepf}, S.~E., \& {Freeman}, K.~C. 1999,
  \aj, 118, 197, \dodoi{10.1086/300919}

\bibitem[{{Kluzniak} \& {Lasota}(2015)}]{kluzniak15}
{Kluzniak}, W., \& {Lasota}, J.~P. 2015, \mnras, 448, L43,
  \dodoi{10.1093/mnrasl/slu200}

\bibitem[{{Lattimer}(2012)}]{lattimer12rev}
{Lattimer}, J.~M. 2012, Annual Review of Nuclear and Particle Science, 62, 485,
  \dodoi{10.1146/annurev-nucl-102711-095018}

\bibitem[{{Lau} {et~al.}(2016){Lau}, {Kasliwal}, {Bond}, {Smith}, {Fox},
  {Carlon}, {Cody}, {Contreras}, {Dykhoff}, {Gehrz}, {Hsiao}, {Jencson},
  {Khan}, {Masci}, {Monard}, {Monson}, {Morrell}, {Phillips}, \&
  {Ressler}}]{lau16}
{Lau}, R.~M., {Kasliwal}, M.~M., {Bond}, H.~E., {et~al.} 2016, \apj, 830, 142,
  \dodoi{10.3847/0004-637X/830/2/142}

\bibitem[{{Lim} {et~al.}(2013){Lim}, {Hwang}, \& {Lee}}]{lim13}
{Lim}, S., {Hwang}, N., \& {Lee}, M.~G. 2013, \apj, 766, 20,
  \dodoi{10.1088/0004-637X/766/1/20}

\bibitem[{{Liu} {et~al.}(2013){Liu}, {Bregman}, {Bai}, {Justham}, \&
  {Crowther}}]{liu13}
{Liu}, J.-F., {Bregman}, J.~N., {Bai}, Y., {Justham}, S., \& {Crowther}, P.
  2013, \nat, 503, 500, \dodoi{10.1038/nature12762}

\bibitem[{{Marchant} {et~al.}(2017){Marchant}, {Langer}, {Podsiadlowski},
  {Tauris}, {de Mink}, {Mandel}, \& {Moriya}}]{marchant17}
{Marchant}, P., {Langer}, N., {Podsiadlowski}, P., {et~al.} 2017, \aap, 604,
  A55, \dodoi{10.1051/0004-6361/201630188}

\bibitem[{{Motch} {et~al.}(2011){Motch}, {Pakull}, {Gris{\'e}}, \&
  {Soria}}]{motch11}
{Motch}, C., {Pakull}, M.~W., {Gris{\'e}}, F., \& {Soria}, R. 2011,
  Astronomische Nachrichten, 332, 367, \dodoi{10.1002/asna.201011501}

\bibitem[{{Motch} {et~al.}(2014){Motch}, {Pakull}, {Soria}, {Gris{\'e}}, \&
  {Pietrzy{\'n}ski}}]{motch14}
{Motch}, C., {Pakull}, M.~W., {Soria}, R., {Gris{\'e}}, F., \&
  {Pietrzy{\'n}ski}, G. 2014, \nat, 514, 198, \dodoi{10.1038/nature13730}

\bibitem[{{Mushtukov} {et~al.}(2015){Mushtukov}, {Suleimanov}, {Tsygankov}, \&
  {Poutanen}}]{mushtukov15}
{Mushtukov}, A.~A., {Suleimanov}, V.~F., {Tsygankov}, S.~S., \& {Poutanen}, J.
  2015, \mnras, 454, 2539, \dodoi{10.1093/mnras/stv2087}

\bibitem[{{Nagao} {et~al.}(2011){Nagao}, {Maiolino}, {Marconi}, \&
  {Matsuhara}}]{nagao11}
{Nagao}, T., {Maiolino}, R., {Marconi}, A., \& {Matsuhara}, H. 2011, \aap, 526,
  A149, \dodoi{10.1051/0004-6361/201015471}

\bibitem[{{Patruno} \& {Zampieri}(2008)}]{patruno08}
{Patruno}, A., \& {Zampieri}, L. 2008, \mnras, 386, 543,
  \dodoi{10.1111/j.1365-2966.2008.13063.x}

\bibitem[{{Patruno} \& {Zampieri}(2010)}]{patruno10}
---. 2010, \mnras, 403, L69, \dodoi{10.1111/j.1745-3933.2010.00817.x}

\bibitem[{{Paxton} {et~al.}(2011){Paxton}, {Bildsten}, {Dotter}, {Herwig},
  {Lesaffre}, \& {Timmes}}]{paxton11}
{Paxton}, B., {Bildsten}, L., {Dotter}, A., {et~al.} 2011, \apjs, 192, 3,
  \dodoi{10.1088/0067-0049/192/1/3}

\bibitem[{{Paxton} {et~al.}(2013){Paxton}, {Cantiello}, {Arras}, {Bildsten},
  {Brown}, {Dotter}, {Mankovich}, {Montgomery}, {Stello}, {Timmes}, \&
  {Townsend}}]{paxton13}
{Paxton}, B., {Cantiello}, M., {Arras}, P., {et~al.} 2013, \apjs, 208, 4,
  \dodoi{10.1088/0067-0049/208/1/4}

\bibitem[{{Paxton} {et~al.}(2015){Paxton}, {Marchant}, {Schwab}, {Bauer},
  {Bildsten}, {Cantiello}, {Dessart}, {Farmer}, {Hu}, {Langer}, {Townsend},
  {Townsley}, \& {Timmes}}]{paxton15}
{Paxton}, B., {Marchant}, P., {Schwab}, J., {et~al.} 2015, \apjs, 220, 15,
  \dodoi{10.1088/0067-0049/220/1/15}

\bibitem[{{Pintore} {et~al.}(2018){Pintore}, {Belfiore}, {Novara},
  {Salvaterra}, {Marelli}, {De Luca}, {Rigoselli}, {Israel}, {Rodriguez},
  {Mereghetti}, {Wolter}, {Walton}, {Fuerst}, {Ambrosi}, {Zampieri}, {Tiengo},
  \& {Salvaggio}}]{pintore18}
{Pintore}, F., {Belfiore}, A., {Novara}, G., {et~al.} 2018, \mnras, 477, L90,
  \dodoi{10.1093/mnrasl/sly048}

\bibitem[{{Rappaport} {et~al.}(2005){Rappaport}, {Podsiadlowski}, \&
  {Pfahl}}]{rappaport05}
{Rappaport}, S.~A., {Podsiadlowski}, P., \& {Pfahl}, E. 2005, \mnras, 356, 401,
  \dodoi{10.1111/j.1365-2966.2004.08489.x}

\bibitem[{{Roberts} {et~al.}(2011){Roberts}, {Gladstone}, {Goulding},
  {Swinbank}, {Ward}, {Goad}, \& {Levan}}]{roberts11}
{Roberts}, T.~P., {Gladstone}, J.~C., {Goulding}, A.~D., {et~al.} 2011,
  Astronomische Nachrichten, 332, 398, \dodoi{10.1002/asna.201011508}

\bibitem[{{Schlegel} {et~al.}(1998){Schlegel}, {Finkbeiner}, \&
  {Davis}}]{schlegel98}
{Schlegel}, D.~J., {Finkbeiner}, D.~P., \& {Davis}, M. 1998, \apj, 500, 525,
  \dodoi{10.1086/305772}

\bibitem[{{Sutton} {et~al.}(2013{\natexlab{a}}){Sutton}, {Roberts},
  {Gladstone}, {Farrell}, {Reilly}, {Goad}, \& {Gehrels}}]{sutton13b}
{Sutton}, A.~D., {Roberts}, T.~P., {Gladstone}, J.~C., {et~al.}
  2013{\natexlab{a}}, \mnras, 434, 1702, \dodoi{10.1093/mnras/stt1133}

\bibitem[{{Sutton} {et~al.}(2013{\natexlab{b}}){Sutton}, {Roberts}, \&
  {Middleton}}]{sutton13}
{Sutton}, A.~D., {Roberts}, T.~P., \& {Middleton}, M.~J. 2013{\natexlab{b}},
  \mnras, 435, 1758, \dodoi{10.1093/mnras/stt1419}

\bibitem[{{Sutton} {et~al.}(2012){Sutton}, {Roberts}, {Walton}, {Gladstone}, \&
  {Scott}}]{sutton12}
{Sutton}, A.~D., {Roberts}, T.~P., {Walton}, D.~J., {Gladstone}, J.~C., \&
  {Scott}, A.~E. 2012, \mnras, 423, 1154,
  \dodoi{10.1111/j.1365-2966.2012.20944.x}

\bibitem[{{Tsygankov} {et~al.}(2017){Tsygankov}, {Doroshenko}, {Lutovinov},
  {Mushtukov}, \& {Poutanen}}]{tsygankov17}
{Tsygankov}, S.~S., {Doroshenko}, V., {Lutovinov}, A.~A., {Mushtukov}, A.~A.,
  \& {Poutanen}, J. 2017, \aap, 605, A39, \dodoi{10.1051/0004-6361/201730553}

\bibitem[{{Tully} {et~al.}(2013){Tully}, {Courtois}, {Dolphin}, {Fisher},
  {H{\'e}raudeau}, {Jacobs}, {Karachentsev}, {Makarov}, {Makarova},
  {Mitronova}, {Rizzi}, {Shaya}, {Sorce}, \& {Wu}}]{2013AJ....146...86T}
{Tully}, R.~B., {Courtois}, H.~M., {Dolphin}, A.~E., {et~al.} 2013, \aj, 146,
  86, \dodoi{10.1088/0004-6256/146/4/86}

\bibitem[{{Villar} {et~al.}(2016){Villar}, {Berger}, {Chornock}, {Margutti},
  {Laskar}, {Brown}, {Blanchard}, {Czekala}, {Lunnan}, \&
  {Reynolds}}]{villar16}
{Villar}, V.~A., {Berger}, E., {Chornock}, R., {et~al.} 2016, \apj, 830, 11,
  \dodoi{10.3847/0004-637X/830/1/11}

\bibitem[{{Voss} {et~al.}(2011){Voss}, {Nielsen}, {Nelemans}, {Fraser}, \&
  {Smartt}}]{voss11}
{Voss}, R., {Nielsen}, M.~T.~B., {Nelemans}, G., {Fraser}, M., \& {Smartt},
  S.~J. 2011, \mnras, 418, L124, \dodoi{10.1111/j.1745-3933.2011.01157.x}

\bibitem[{{Walton} {et~al.}(2015){Walton}, {Harrison}, {Bachetti}, {Barret},
  {Boggs}, {Christensen}, {Craig}, {Fuerst}, {Grefenstette}, {Hailey},
  {Madsen}, {Middleton}, {Rana}, {Roberts}, {Stern}, {Sutton}, {Webb}, \&
  {Zhang}}]{walton15b}
{Walton}, D.~J., {Harrison}, F.~A., {Bachetti}, M., {et~al.} 2015, \apj, 799,
  122, \dodoi{10.1088/0004-637X/799/2/122}

\bibitem[{{Walton} {et~al.}(2018){Walton}, {F{\"u}rst}, {Heida}, {Harrison},
  {Barret}, {Stern}, {Bachetti}, {Brightman}, {Fabian}, \&
  {Middleton}}]{walton18}
{Walton}, D.~J., {F{\"u}rst}, F., {Heida}, M., {et~al.} 2018, \apj, 856, 128,
  \dodoi{10.3847/1538-4357/aab610}

\bibitem[{{Wang} {et~al.}(2015){Wang}, {Liu}, {Bai}, \& {Guo}}]{wang15}
{Wang}, S., {Liu}, J., {Bai}, Y., \& {Guo}, J. 2015, \apjl, 812, L34,
  \dodoi{10.1088/2041-8205/812/2/L34}

\bibitem[{{Wiktorowicz} {et~al.}(2017){Wiktorowicz}, {Sobolewska}, {Lasota}, \&
  {Belczynski}}]{wiktorowicz17}
{Wiktorowicz}, G., {Sobolewska}, M., {Lasota}, J.-P., \& {Belczynski}, K. 2017,
  \apj, 846, 17, \dodoi{10.3847/1538-4357/aa821d}

\bibitem[{{Wilson-Hodge} {et~al.}(2018){Wilson-Hodge}, {Malacaria}, {Jenke},
  {Jaisawal}, {Kerr}, {Wolff}, {Arzoumanian}, {Chakrabarty}, {Doty},
  {Gendreau}, {Guillot}, {Ho}, {LaMarr}, {Markwardt}, {{\"O}zel}, {Prigozhin},
  {Ray}, {Ramos-Lerate}, {Remillard}, {Strohmayer}, {Vezie}, {Wood}, \& {NICER
  Science Team}}]{wilsonhodge18}
{Wilson-Hodge}, C.~A., {Malacaria}, C., {Jenke}, P.~A., {et~al.} 2018, \apj,
  863, 9, \dodoi{10.3847/1538-4357/aace60}

\end{thebibliography}



\end{document}